\documentclass[11pt,a4paper]{article}

\usepackage[margin=1in]{geometry}
\usepackage[T1]{fontenc}
\usepackage{lmodern}
\usepackage{microtype}
\usepackage{setspace}

\usepackage{amsmath}
\usepackage{amssymb}
\usepackage{amsfonts}
\usepackage{bm}
\usepackage{siunitx}

\usepackage{graphicx}
\usepackage{booktabs}
\usepackage{multirow}
\usepackage{array}
\usepackage{float}
\usepackage{subcaption}

\usepackage[authoryear,round]{natbib}
\usepackage[hidelinks]{hyperref}
\usepackage{url}

\usepackage{enumitem}
\usepackage[dvipsnames]{xcolor}

\newcommand{\bipin}[1]{\textcolor{blue}{#1}}

\usepackage{authblk}

\title{
	{\normalfont\itshape
		\bipin{Preprint submitted for peer review}}
	\\[0.8cm]
	\bfseries
	PrecipCorrDiff: A Deep Learning Framework for Generating 0.0625$^\circ$ High-Resolution IMD Gridded
	Precipitation Data
}

\author[1]{Bipin Kumar\thanks{
		Corresponding author:
		\texttt{bipink@tropmet.res.in}}}

\author[1]{Rajib Chattopadhyay}
\author[2]{Harsh P. Bachhav}
\author[2]{Jannu}
\author[1]{Bhupendra Bahadur Singh}
\author[3]{Ramesh Ragala}
\author[3]{Ananda Kumar Das}
\author[4]{Manish Modani}
\author[1]{Thara Prabhakaran}
\author[3]{Mrutyunjay Mohapatra}

\affil[1]{
	Indian Institute of Tropical Meteorology,
	Ministry of Earth Sciences,
	Government of India,
	Pune, India
}

\affil[2]{
	Vishwakarma Institute of Technology,
	Bibwewadi, Pune, India
}

\affil[3]{
	India Meteorological Department,
	Ministry of Earth Sciences,
	Government of India,
	Lodhi Road, New Delhi, India
}

\affil[4]{
	NVIDIA Graphics India Pvt. Ltd.,
	Mahadevapura, Bengaluru, Karnataka, India
}

\date{}

\begin{document}

	\maketitle

%
%
\begin{abstract}

High-resolution precipitation datasets are essential for weather forecasting, hydrological modeling, and climate risk assessment, yet observation-based gridded products often lack the spatial detail required for regional applications. Here we present PrecipCorrDiff, a diffusion-based downscaling framework built by modifying the original corrector diffusion based archetecture that incorporates a customized loss function and a multi-scale residual generation strategy to improve the precipitation especially the representation of extremes. PrecipCorrDiff generates 0.0625° resolution India Meteorological Department (IMD) precipitation fields from 0.25° gridded data while preserving precipitation statistics and reconstructing fine-scale spatial variability. Validation against independent IMD station observations, comparing the proposed framework with the SRGAN-based model and the original CorrDiff model, indicates that the proposed framework substantially improves the representation of spatial precipitation patterns, rainfall intensity distributions, and extreme rainfall events, while eliminating the systematic negative bias present in the previously developed SRGAN-based model. Across four climatologically distinct homogeneous Indian monsoon regions, PrecipCorrDiff reproduces observed precipitation characteristics with near-zero spatial bias, demonstrating the potential of diffusion-based downscaling for generating physically consistent, high-resolution precipitation datasets for hydrological, meteorological, and climate applications.
\end{abstract}

\section{Introduction}

High-resolution precipitation information underpins hydrological modelling, flood forecasting, weather nowcasting, disaster risk management, agricultural planning, and water resource operations, yet global and regional models, together with many gridded observational products, remain too coarse to resolve local rainfall variability \citep{Mardani2025, Srivastava2024, Wang2021}. This scale mismatch is particularly acute for precipitation, whose intensity and spatial structure vary sharply over short distances and are strongly modulated by topography, coastlines, and multiscale atmospheric processes \citep{Srivastava2024,Harris2022, Kumar2021,Kumar2026LocationSpecific}.

Over the Indian region, the India Meteorological Department (IMD) gridded rainfall dataset, constructed from a dense rain-gauge network, is the most widely used precipitation product for monsoon monitoring, drought and flood assessment, and model validation \citep{Pai2014, Rajeevan2006}. However, its standard resolutions ($1.0^o$ and $0.25^o$) are insufficient for catchment-scale hydrological and agricultural applications, motivating sustained efforts to downscale IMD rainfall to finer resolutions while preserving the statistical fidelity of the original record.  Recently, Peringiyil et al. (2025) introduced the Indian Precipitation Ensemble Dataset (IPED), a station-based 0.1° daily gridded ensemble product for India that addresses key limitations of deterministic grids like IMD's. IPED comprises a 30-member daily ensemble at both $0.1^o$ and $0.25^o$ resolutions for 1991–2020, generated using a locally weighted spatial regression approach on the largest available rain gauge network. This method explicitly incorporates topographical predictors (elevation, slope, aspect), yielding superior discrimination and reliability across all thresholds compared to IMD's Shepard interpolation. IPED's probabilistic nature enhances utility for data assimilation, hydrologic modelling, and extreme event validation, though its recent temporal span limits long-term climate studies. Still, has limitations for usage at block level.

Dynamical and statistical downscaling approaches remain widely used but are often computationally expensive, limited in spatial detail, or constrained in representing nonlinear precipitation behaviour and extremes \citep{Srivastava2024, Liu2025,Golla2024}. Deep learning has consequently emerged as a practical alternative, learning direct mappings from coarse to fine-scale precipitation and frequently outperforming traditional methods in reconstructing local spatial features \citep{Sun2024,Kumar2023,Wang2021}. CNN-based super-resolution models, residual networks, ConvLSTM, U-Net, GANs, and, more recently, diffusion models have all been explored for this task\citep{Sun2024,Kumar2023, Golla2024}, with CNN and residual-learning approaches showing strong skill in reproducing spatial and temporal rainfall patterns and, in some settings, extremes \citep{Wang2021,Golla2024}.

For the Indian region, deep-learning downscaling of IMD gridded rainfall has an established history. Early stacked-CNN and DeepSD-based approaches improved correlation and error relative to interpolation and helped preserve spatial discontinuities and intensity gradients over the monsoon region \citep{Kumar2021,Harilal2021}. Subsequent comparisons across DeepSD, ConvLSTM, U-Net, and SRGAN found that SRGAN outperformed the other architectures overall, largely because its perceptual, adversarial loss reduced the over-smoothing associated with MSE-driven training and better captured fine spatial detail \citep{Kumar2023}. Related studies similarly reported that SRGAN and residual-learning approaches produce more perceptually realistic rainfall maps than conventional benchmarks such as Bias Correction and Spatial Disaggregation (BCSD), with gains in distributional structure, mean behaviour, and extremes \citep{Murukesh2023}, while residual networks have performed well for IMD downscaling from $1.0^o$ to $0.25^o$ \citep{Golla2024}.  Even so, studies focused on Indian monsoon subseasonal variability indicate that standard super-resolution architectures can still struggle with extreme rainfall unless loss functions or architectures are explicitly modified to emphasize these aspects \citep{Chandel2025}. Weighted loss functions have emerged as an effective, architecture-agnostic strategy for this purpose. \cite{Chandel2025}  introduced a weighted MSE loss (YNet\_DE) that up-weights values below the $10^{th}$ and above the $90^{th}$ percentile, reducing bias in extreme indices such as $R95 \,\,mm$ and $R20\,\, mm$ relative to the unweighted YNet and the dynamically-informed YNet\_D, with improved transferability to GCM outputs. Similarly, \cite{Murukesh2023}  showed that a weighted MAE loss (SRDRN-WMAE), scaled by precipitation intensity, achieved the highest quantile skill scores at the $90^{th}–99^{th}$ percentiles and the lowest systematic bias among tested configurations, while also suppressing checkerboard artifacts common to CNN-based super-resolution.

However, both studies report trade-offs. YNet\_DE showed slightly higher bias in mean daily rainfall and reduced skill in intraseasonal active/break composites compared to YNet\_D, while SRDRN-WMAE substantially overestimated wet-spell persistence (consecutive wet days) despite its gains in intensity-based indices. These findings suggest that weighted loss functions robustly enhance extreme precipitation representation, but the weighting scheme requires careful calibration to avoid degrading mean-state accuracy or event-persistence characteristics. The above points are directly relevant to the present study. In our own prior SRGAN-based downscaling of the IMD gridded dataset, the model produced visually realistic high-resolution rainfall fields but systematically underestimated the mean precipitation distribution when evaluated against the complete IMD dataset and independent ground-based rain-gauge observations (refer Fig. 2a). Because climate and hydrological applications depend not only on plausible-looking rainfall maps but on downscaled fields whose mean, intensity distribution, and spatial variability faithfully match observations, the bias represents is an important limitation of GAN-based downscaling that must be addressed before such methods can be reliably applied operationally.

Diffusion probabilistic models offer a promising route to overcoming the limitation described above. Rather than learning a single deterministic or adversarially optimized mapping, diffusion models represent the full conditional distribution of plausible high-resolution precipitation fields given coarse input (refer figure 3), which is important for unbiased means, realistic variability, and extreme-event behaviour \citep{Harris2022, Srivastava2024, Schmidt2025}. Diffusion-based downscaling has been shown to sharpen coherent spatial structures, recover realistic spectra and distributions, and preserve large-scale precipitation while improving fine-scale organization \cite{Mardani2025,Aich2026}, and to outperform strong deterministic and generative baselines, particularly for high-frequency structure and extremes \citep{Srivastava2024}. Wasserstein-regularized diffusion has reduced empirical bias at extreme intensities \citep{Liu2025}, and conditional diffusion frameworks have shown strong performance for bias correction and small-scale structure recovery while preserving physically coherent behaviour \citep{Shidqi2023, Aich2026, Lee2026}. Benchmarking evidence further indicates that diffusion models are more stable to train than GANs and better recover sub-grid variability and heavy-tailed distributions that deterministic and adversarial methods tend to smooth or suppress \citep{Lee2026}, with competitive or superior geographic generalization relative to GANs (Harder et al., 2026).    Other advanced architectures, such as Fourier Neural Operators and Graph Neural Operators, have also been applied to climate downscaling \citep{Yang2023, Jiang2023, Blasone2025}, but, to our knowledge, have not yet been evaluated for Indian or IMD precipitation data. Indian precipitation presents a particularly challenging test case for generative downscaling because it exhibits pronounced spatial and temporal heterogeneity arising from the southwest and northeast monsoons, complex topography, coastal influences, and localized convective systems. These characteristics produce highly multimodal rainfall distributions and sharp spatial gradients, suggesting that diffusion-based generative models, which learn complex probability distributions rather than deterministic mappings, may be particularly well suited to reconstructing realistic fine-scale precipitation patterns and extremes. We investigate this hypothesis using the IMD gridded precipitation dataset.

Present study develops a diffusion-model-based framework for downscaling the IMD gridded precipitation dataset, designed to overcome the mean-precipitation underestimation and related statistical shortcomings observed in previous SRGAN-based approach while retaining realistic fine-scale spatial patterns. Specifically, this study aims to: (i) develop a diffusion-probabilistic-model framework for downscaling IMD gridded rainfall and (ii) a comparison focused on reproduction of the mean, full intensity distribution, and spatial variability against both the IMD gridded dataset and the independent gauge observations. To our knowledge, this constitutes the first application of a diffusion-based downscaling framework to the IMD gridded precipitation dataset, providing a direct, statistically grounded comparison against a GAN-based baseline and contributing a South Asian case study to the growing diffusion-based precipitation downscaling literature.

\section{Data sources and pre-processing }
\begin{figure}
	\center
	\includegraphics[width=0.9\textwidth,height=0.3\textheight]{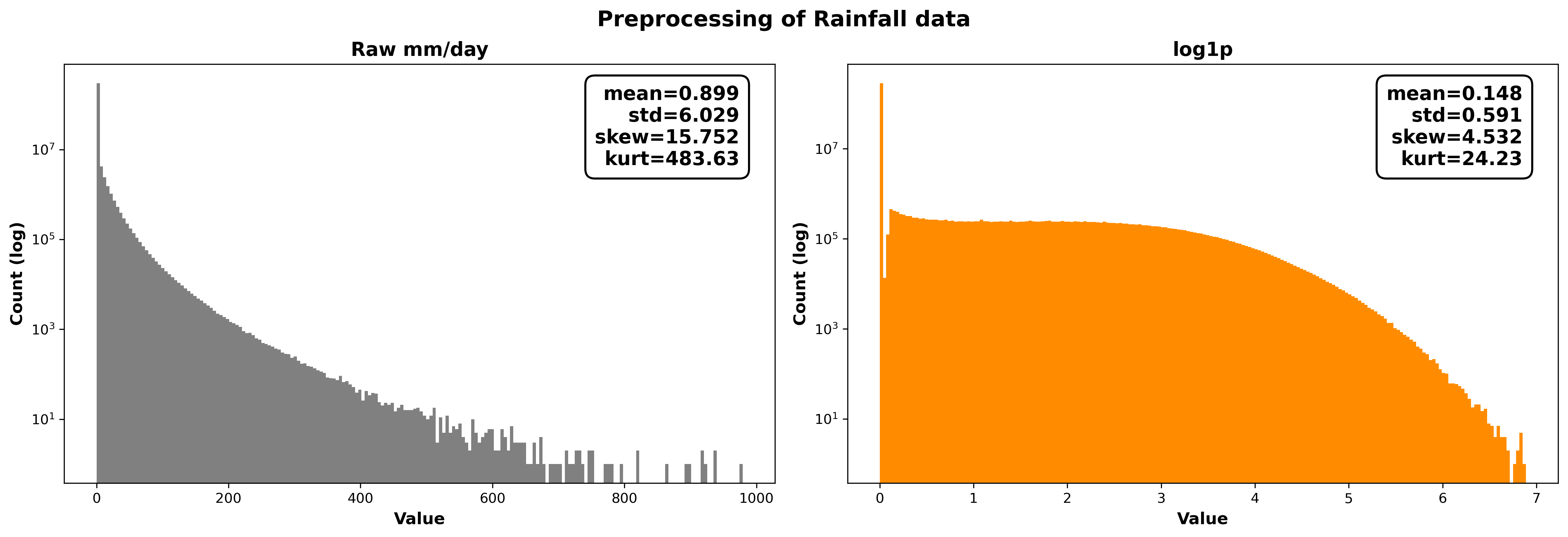}
	\caption{Representation of Log (1+p) transformation used to remove the skewness of the data}
	\label{fig:log1p}
\end{figure}

The primary dataset used in this study is the daily gridded rainfall product developed by the India Meteorological Department (IMD). This dataset is a well-established and high-quality observational product with a spatial resolution of $0.25^o \times 0.25^o$ (approximately 25 km × 25 km). For the present analysis, a continuous 49-year record (1975–2023) covering the entire Indian subcontinent was utilized. The availability of such a long-term dataset is crucial for capturing climatic variability and ensuring reliable downscaling experiments.  Rainfall distribution over India is strongly influenced by complex topography, particularly in regions such as the Himalayas, the Western Ghats, and the northeastern hill ranges. The other variables are sourced from the ERA5 reanalysis product, which provides atmospheric state variables including two-meter dew point temperature (D2M) and a static orography file.  The preprocessing pipeline applies a log-plus-one transformation to all precipitation values (P), i.e., log(1 + P), to reduce the heavy right-skew inherent in rainfall distributions, as represented in the figure \ref{fig:log1p}, and to stabilize variance during training. This transformation maps dry days (P = 0) to zero and ensures that gradient flow is meaningful across the full dynamic range of observed rainfall. 

The dataset is partitioned chronologically into year-wise training (80\%), validation, and testing (20\%) subsets. Chronological splitting is deliberately chosen over random shuffling to prevent temporal data leakage, as atmospheric fields exhibit strong serial autocorrelation. A temporal conditioning buffer of T=5 preceding time steps is maintained for each sample to provide the model with short-term memory of evolving precipitation patterns. To evaluate the robustness of the proposed framework under diverse climatic conditions, model performance is further assessed across four homogeneous rainfall regions of India: Northern India, Central India, South Peninsular India, and North Eastern India. These regions represent distinct monsoon rainfall characteristics and provide a comprehensive assessment of the model's ability to generalize across different precipitation regimes. In addition, model performance is evaluated across multiple rainfall intensity categories using predefined precipitation thresholds to examine its capability in reproducing light, moderate, heavy, and extreme rainfall events.

\section{Model Development}

\subsection{Baseline model}

We first extended the SRGAN-based downscaling method previously developed by Kumar et al. (2023) by incorporating several additional input variables and systematically evaluating their impact on model performance. Among the configurations tested, the best-performing model was obtained by adding just two extra input variables, dew point temperature (D2m) and data variance (V), to the original SRGAN formulation. This configuration was used to generate downscaled data at 0.0625° resolution. This modified model is used as baseline for performance evaluation of the proposed PrecipCorrDiff.

The downscaled output was validated against the original IMD gridded dataset and independent station observations for the test period, using two complementary diagnostics: the probability density function (PDF) of the data distribution, and the rainfall distribution for the test data for JJAS (June–September) period. Figure 2 presents this comparison and shows that the mean and actual rainfall values estimated by SRGAN are significantly consistently underestimated.

\begin{figure}[ht]
	\center
	\includegraphics[width=0.99\textwidth,height=0.27\textheight]{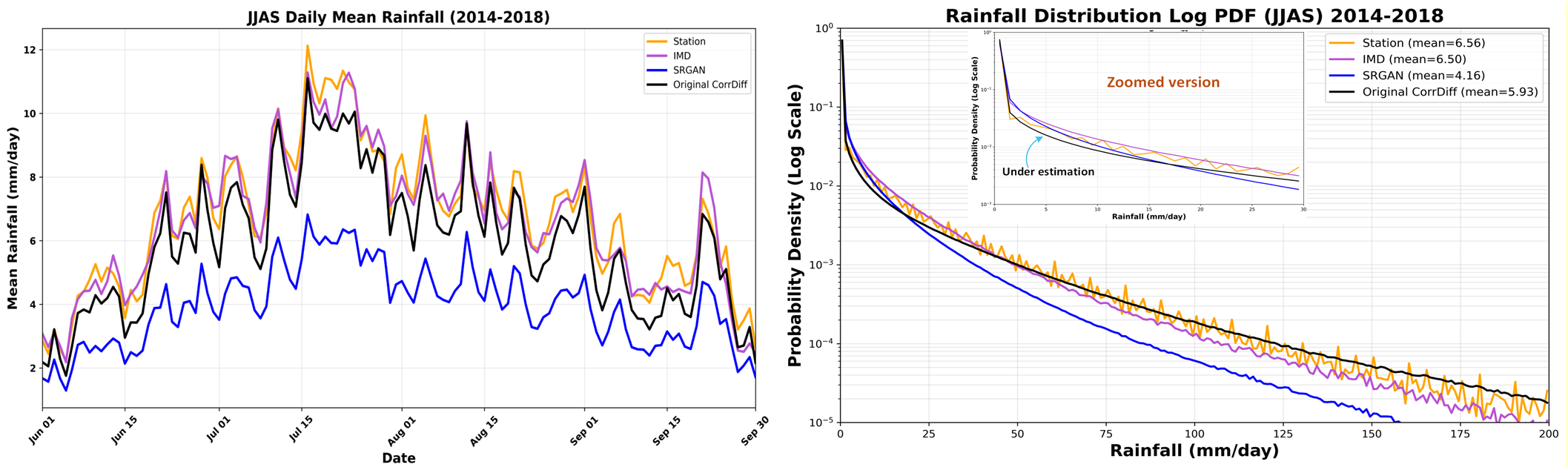}
	\caption{Comparison of rainfall representations from the SRGAN model output, original IMD data, and station observations. The left panel shows the mean rainfall density for JJAS months over the 2014–2018 period, while the right panel presents the corresponding raw data frequencies for the same period.}
	\label{fig:srgan_underpredicted}
\end{figure}

The mean rainfall values shown in Figure \ref{fig:srgan_underpredicted}(a) were computed by first averaging the rainfall for each calendar day (e.g., 1 June) across the five-year study period (2014–2018) at every station (or grid point). The resulting values were then spatially averaged over all station locations for the station dataset and over all grid points for the IMD and SRGAN datasets, yielding a single regional mean rainfall value for each day

This systematic underestimation exposed a fundamental limitation of SRGAN: its inability to reliably capture long-term rainfall values, which made it unsuitable for generating dependable long-term, high-resolution precipitation datasets. It was precisely this persistent under-prediction that pushed us to look beyond SRGAN and seek an alternative downscaling approach that could overcome this shortcoming. 

\begin{figure}[ht]
	\center
	\includegraphics[width=0.9\textwidth,height=0.35\textheight]{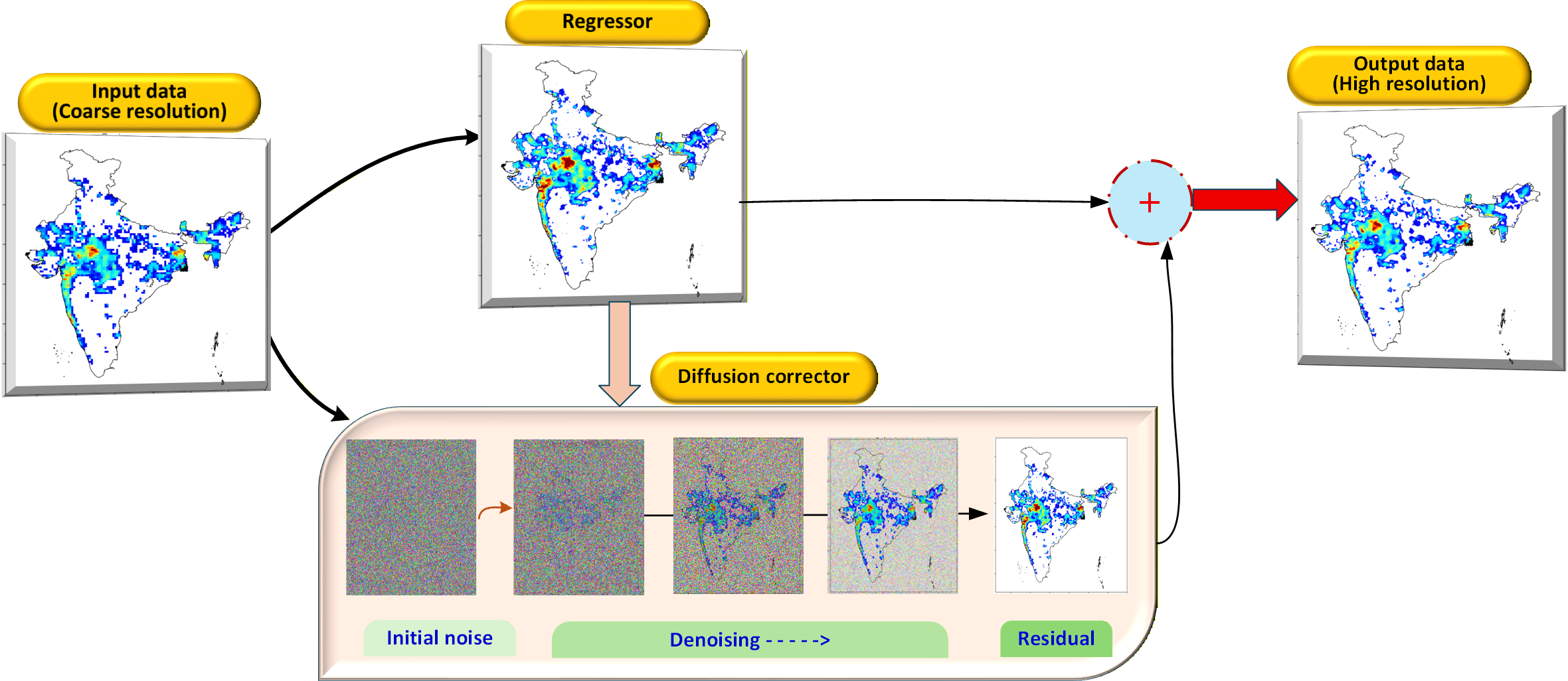}
	\caption{Representation of Corrector Diffusion model for data downscaling.}
	\label{fig:intro_diffusion}
\end{figure}

Recent advances in diffusion models have demonstrated their effectiveness in generative modeling and geophysical super-resolution, making them well suited for precipitation downscaling \cite{Mardani2025, Srivastava2024}. Among these approaches, the Corrector Diffusion (CorrDiff) framework provides a computationally efficient two-stage generative architecture that combines deterministic regression with stochastic residual correction, as illustrated in Figure \ref{fig:intro_diffusion}. We first applied the original CorrDiff framework, referred to hereafter as  CorrDiff, to our precipitation dataset and performed the same comparison presented in Figure \ref{fig:srgan_underpredicted}. Although CorrDiff improved the representation of the mean rainfall compared with SRGAN, it continued to underestimate rainfall at the lower end of the precipitation range. This limitation motivated us to further extend the CorrDiff architecture to better represent the full precipitation distribution.

To mitigate this underestimation, architectural improvements to CorrDiff were incorporated. A key challenge inherent to rainfall data is its pronounced class imbalance, characterized by a high frequency of low-rainfall samples relative to a comparatively sparse representation of high-rainfall events. This multiscale characteristic of the data distribution is likely a contributing factor to the underestimation bias observed in the annual cycles (Figure \ref{fig:srgan_underpredicted}). Architectural mechanisms such as multi-head attention and dilated bottleneck structures within the UNet framework have been shown to improve model performance under such data distributions \citep{Juhong2024, Chen2017, Wang2022}. 

Building on this paradigm, we develop the proposed PrecipCorrDiff framework for high-resolution precipitation downscaling, which is described in detail in the following subsection.

\subsection{\textbf{Proposed PrecipCorrDiff Framework}}

The proposed PrecipCorrDiff (Precipitation Corrector Diffusion Framework) is built upon the Corrector Diffusion (CorrDiff) paradigm for precipitation downscaling proposed by NIVIDA \citep{Mardani2025}. While retaining the two-stage generative formulation of CorrDiff, PrecipCorrDiff incorporates a customized loss function and a modified residual generation strategy to improve the reconstruction of precipitation extremes. Specifically, the framework decomposes the downscaling task into two sequential stages: (i) a deterministic regression model that predicts the large-scale precipitation field , and (ii) a stochastic diffusion-based corrector that reconstructs the fine-scale residual structures. The two-stage design is motivated by the known shortcomings of purely deterministic or purely generative downscaling models. Deterministic models tend to over-smooth small-scale spatial variability, producing blurry precipitation fields that suppress extreme events. Purely stochastic models, on the other hand, may generate physically plausible textures without maintaining correct spatial mean states. The CorrDiff framework resolves this by first obtaining a high-quality deterministic mean field from the regression stage (Stage 1: refer Figure 4) and then employing a conditional diffusion model (Stage 2: refer right part of Figure 5) to correct the residuals using UNet architecture, injecting physically consistent fine-scale variance around the mean  \citep{Mardani2025}. 

\begin{figure}[ht]
	\center
	\includegraphics[width=0.7\textwidth,height=0.4\textheight]{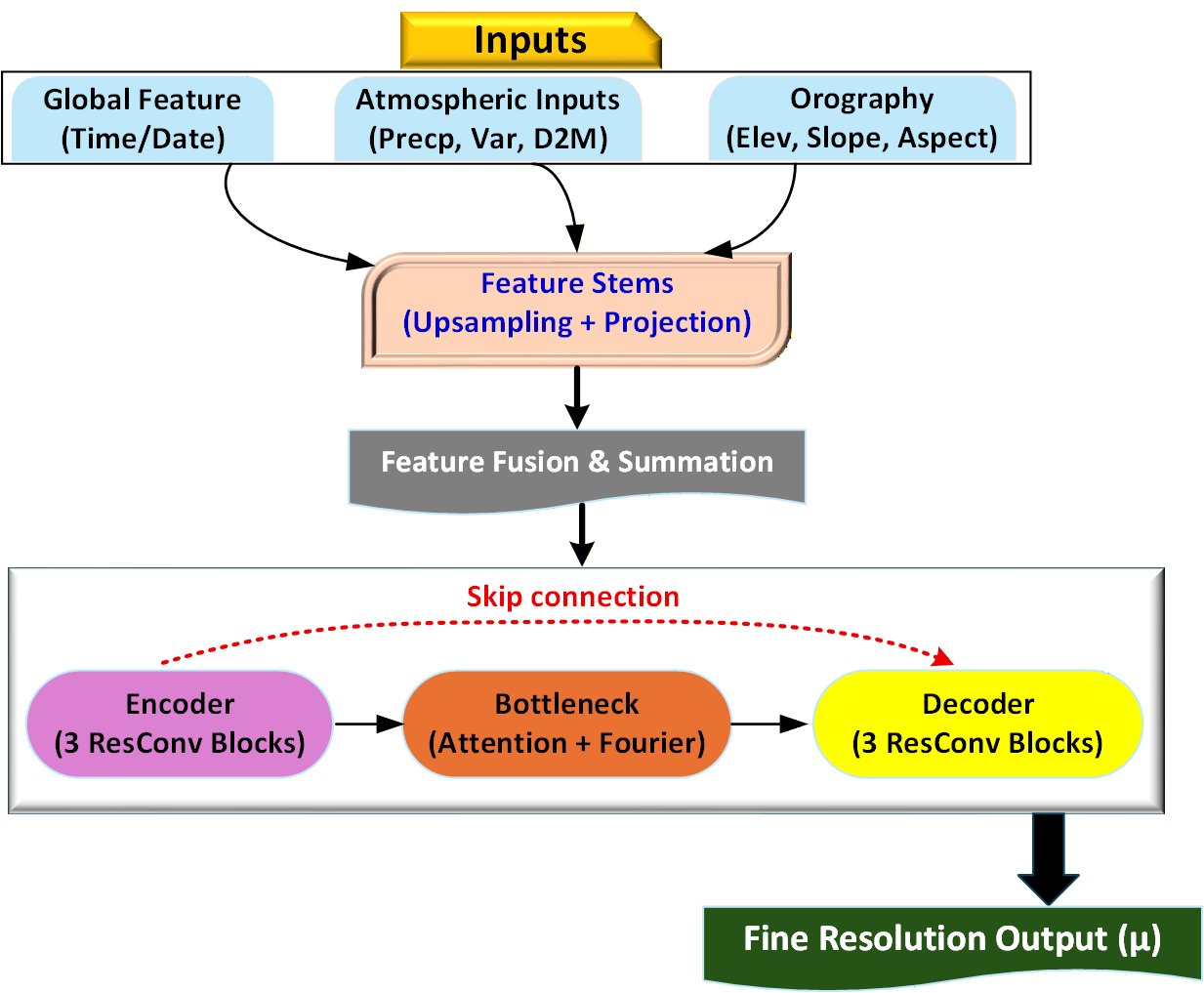}
	\caption{Representation of Stage 1 of the proposed Corrector Diffusion model.}
	\label{fig:regressor}
\end{figure}

Coarse-resolution inputs for the UNet are derived during training by applying average pooling with a kernel size equal to the downscaling factor (DS\_FACTOR = 4) to the fine-resolution target field, thereby maintaining exact consistency between input and target grids across all training samples. The overall predicted precipitation field is thus the sum of the deterministic mean and the learned stochastic residual. The details of modified stages for the proposed PrecipCorrDiff are provided in this section.

\begin{figure}[htpb]
	\center
	\includegraphics[width=0.99\textwidth,height=0.85\textheight]{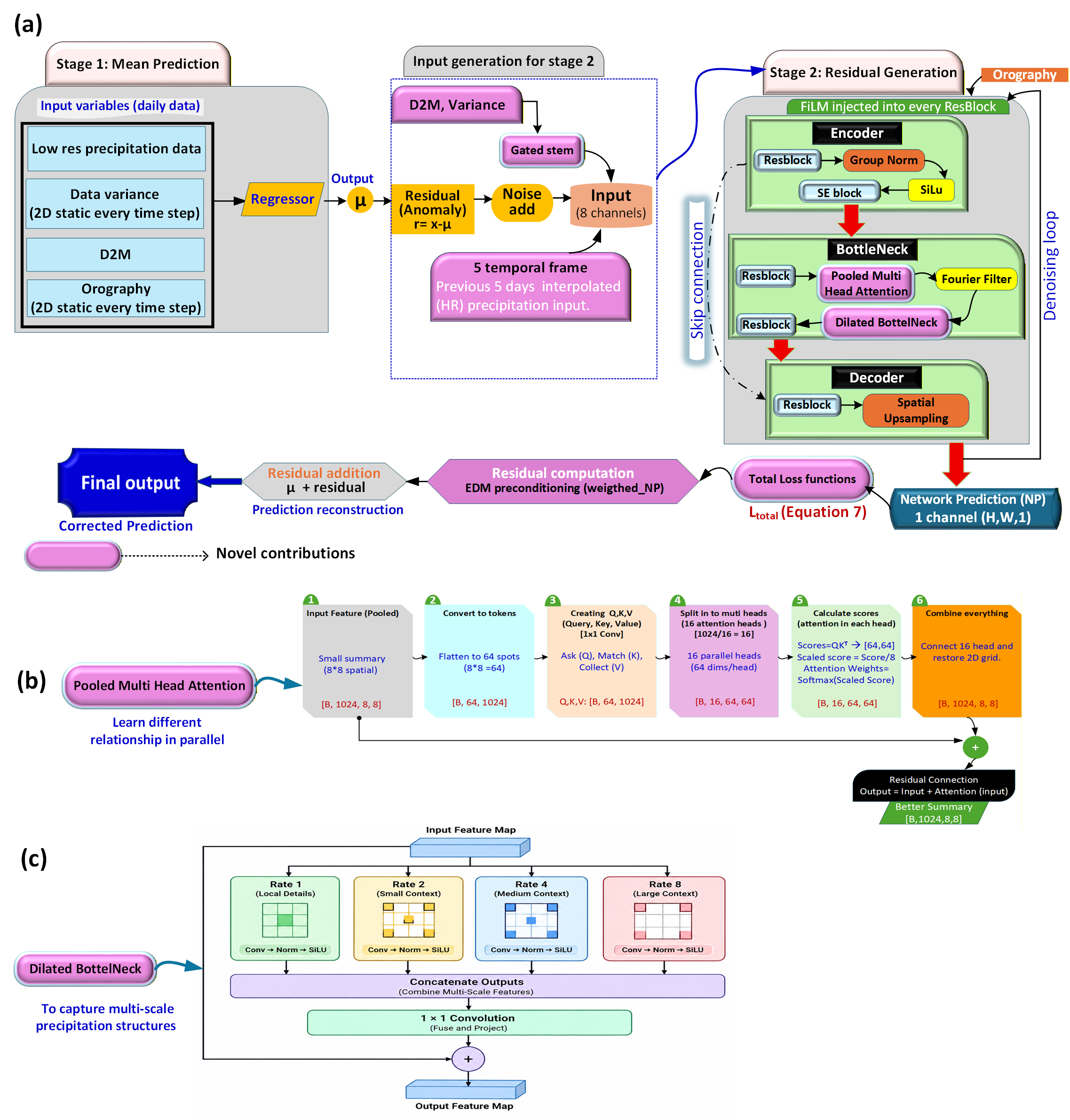}
	\caption{Overview of the developed PrecipCorrDiff architecture for precipitation downscaling. (a) Overall model architecture, where the pink blocks indicate the newly introduced components relative to the original CorrDiff framework. (b) Pooled multi-head attention module used to learn spectral dependencies through Fourier features. (c) Dilated bottleneck module with dilation rates of {1, 2, 4, 8} for capturing precipitation features at multiple spatial scales.}
	\label{fig:full_architecture}
\end{figure}

\subsubsection{Proposed Deterministic Regression Network (Stage 1)   }
The Stage 1 component is a fully convolutional encoder-decoder network that produces a deterministic estimate of the fine-resolution precipitation field from the coarse-resolution input and auxiliary conditioning fields. It uses dual parallel encoder branches: an $r-stem$ that concatenates the coarse precipitation and variance map, upsamples them via bilinear interpolation, and applies a coordinate-aware convolutional projection; and a $t-stem$ that independently encodes the three-channel topographic field. Both streams pass through cascaded residual blocks at progressively coarser scales, with skip connections linking corresponding encoder and decoder levels, as illustrated in Figure \ref{fig:regressor}. The D2M field enters through a separate $d-stem$ and is interpolated to fine resolution before being added to the main feature stream, preserving its distinct physical meaning from the rainfall predictor.

Global scalar features (day-of-year, hour-of-day) pass through a learned embedding MLP and are broadcast spatially into each branch. The regressor is trained independently and frozen during Stage 1 (joint fine-tuning remains unexplored), so its output serves as the deterministic mean field around which the Stage 2 of the model learns physically consistent residuals decoupling that prevents the diffusion objective from destabilizing the mean estimator and allows the regressor to be replaced or fine-tuned independently.

\subsubsection{Proposed Unet-based Residual Generation (Stage 2)}

The Stage 2 network is a UNet-based conditional denoising model that predicts the residual field, defined as the difference between the fine-resolution target and the frozen Stage 1 mean prediction, using the EDM (Elucidating Diffusion Models) preconditioning framework. The denoising function is parameterized through analytically derived input, output, and skip-scaling coefficients as functions of the noise level $\sigma$ , computed from a dataset-specific $\sigma_{data}$ (0.1925 for current test data set 2014-2023) estimated from the standard deviation of the log-transformed residuals. The network receives, as input, the concatenation of noisy residual, the Stage 1 mean prediction, and five temporally conditioned frames upsampled to fine resolution . Additional conditioning is provided by a fine-resolution D2M field and a coarse-resolution climatological variance map, each processed through dedicated gated stems with zero-initialized learnable gates to preserve identity at initialization and enable seamless transfer from pretrained checkpoints as depicted in middle part of the Figure  \ref{fig:full_architecture}(a).  The topographic information is injected at every resolution through FiLM (Feature-wise Linear Modulation) conditioning layers (refer last part of Figure  \ref{fig:full_architecture}(a))
To prevent the terrain data from introducing dangerous numerical values, we wrapped the injection in tanh gated activations, while Squeeze-and-Excitation (SE) blocks throughout the network adaptively re-weight channel contributions. 
 The encoder comprises four resolution levels with channel multipliers of (1,2,2,4) relative to a base width of 256, each containing three residual blocks followed by stride down sampling. Every residual block incorporates GroupNorm, SiLU activation, noise-level embedding, FiLM modulation, dropout (p=0.01)), and squeeze-and-excitation recalibration. 
 At this stage, we introduce the bottleneck, multi-head bottleneck attention (BnAttn) which operates on a spatially subsampled feature map to capture long-range dependencies at reduced memory cost. The bottleneck integrates pooled multi-head attention (for detail, refer Figure \ref{fig:full_architecture}(b)) , a learnable Fourier feature filter for spectral dependency modeling, and an Atrous Spatial Pyramid Pooling (ASPP), also known as Dialted bottleneck, module with dilation rates ({1,2,4,8}) to capture multi-scale precipitation structures (refer Figure \ref{fig:full_architecture}(c)). The decoder symmetrically mirrors the encoder with skip connections, enabling effective reconstruction of high-resolution residual fields.

The overall architecture of developed corrector  diffusion UNet  framework (PrecipCorrDiff) is illustrated in Figure \ref{fig:full_architecture} (a), highlighting the hybrid two-stage framework and the interaction between the deterministic Stage 1 prediction and the probabilistic Stage 2 residual correction process. The figure clearly depicts the sequence of operations in both stages and their interconnections within the complete model pipeline. The principal architectural innovations over existing Corrector diffusion model are as follows: 

\begin{enumerate}	[itemsep=-1mm] 
	\item [(i)] Implementation of Dynamic Topography Injection via FiLM.
	\item[(ii)] The introduction of the enhanced bottleneck module, which combines Fourier filtering and ASPP for improved multi-scale feature extraction. 
	\item[(iii)] The incorporation of multi-head pooled attention to effectively capture long-range spatial dependencies while maintaining computational efficiency and
	\item[(iv)] Development of physics-based loss function.
\end{enumerate}

\subsection{\textbf{Loss function development}}
Stage 2 training employs a physics-informed composite loss (hybrid\_sigma\_loss) combining five complementary terms. The primary spatial term is a noise-level-weighted mean absolute error with intensity-based sample weighting to emphasize heavy rainfall,

\begin{equation}
	L_{\mathrm{spatial}} =
	\mathbb{E}\left[
	\left|\widehat{x}_{0} - x_{0}\right|  	\cdot  	\left(\max(x_{0},0)^{\alpha} + 1.0\right)
	\cdot   	\min(\mathrm{SNR},5.0)  	\right]
	\label{eq:spatial_loss}
\end{equation}

Where, $x_{0}$ represents the ground-truth value and $\widehat{x}_{0}$ represents the predicted value. The parameter $\alpha = 1.5$ controls the exponential emphasis on extreme precipitation values, while the minimum Signal-to-Noise Ratio (SNR) weighting,

\begin{equation}
	\mathrm{SNR} = \frac{\sigma_{\mathrm{data}}^{2}}{\sigma_{l}^{2}},
\end{equation}

follows the EDM practice to up-weight loss contributions at moderate noise levels. The $+1.0$ offset ensures that dry-background pixels still receive standard MAE gradients. A phase-alignment term based on the log-cosh of the Pearson correlation coefficient (PCC) error is applied to overcome saturation near correlation plateaux:

\begin{equation}
	L_{\mathrm{PCC}} =  \log\left[ 	\cosh\left(5.0 \cdot (1-\mathrm{PCC})\right) 	\right]  +
	0.1\left|  	\log(\mathrm{var}_{\mathrm{pred}}) 	-  	\log(\mathrm{var}_{\mathrm{target}}) \right|
	\label{eq:pcc_loss}
\end{equation}

which steepens the gradient in the range $\mathrm{PCC} \in [0.8, 1.0]$, where standard MSE-based losses lose discriminative power, and adds a variance-mismatch penalty between the predicted and target fields. This term is delayed for the first 10 training epochs, with $\lambda_{\mathrm{PCC}} = 0$ for epochs $< 10$, so the model first learns the structural mean before aligning fine-scale phase structure, and its weight then grows linearly to a maximum of $\lambda_{\mathrm{PCC}} = 6.0$. A spectral consistency term computes the mean absolute difference between the 2D real FFT (rfft2) magnitude spectra of the predicted and target residuals, ensuring the model reproduces the correct spatial frequency distribution and does not suppress high-wavenumber variability,

\begin{equation}
	L_{\mathrm{spectral}} =
	\mathbb{E}\left[
	\left\|
	F\left(\widehat{x}_{0}\right) -
	F\left(x_{0}\right)
	\right\|
	\right]
	\label{eq:spectral_loss}
\end{equation}

while a soft dry-day penalty discourages false precipitation on genuinely dry grid cells---pixels where the target value falls below a trace threshold of $10^{-4}$ in log space---by rectifying any predicted value exceeding $0.1$ mm,

\begin{equation}
	L_{\mathrm{dry}} =
	\mathbb{E}\left[
	\operatorname{ReLU}\left(\widehat{x}_{0} - 0.1\right)
	\cdot
	\mathbb{I}\left(x_{0} \leq 10^{-4}\right)
	\right]
	\label{eq:dry_loss}
\end{equation}

an asymmetric formulation that avoids the gradient instability of hard masking while discouraging spurious light drizzle across dry regions. In Eq.~\ref{eq:dry_loss}, $\mathbb{I}$ denotes the indicator function, which behaves as a step function based on the given threshold value. Finally, a mass conservation term penalizes systematic bias in the domain-total spatial mean of predicted precipitation relative to the target,

\begin{equation}
	L_{\mathrm{mass}} = \mathbb{E}\left[	\left| 	\widehat{x}_{0} - \left\langle x_{0} \right\rangle
	\right|  	\right]
	\label{eq:mass_loss}
\end{equation}

giving the total loss

\begin{equation}
	L_{\mathrm{total}} = L_{\mathrm{spatial}} + 0.10L_{\mathrm{spectral}} +\lambda_{\mathrm{PCC}}L_{\mathrm{PCC}} + 0.05L_{\mathrm{dry}} 	+ 0.20L_{\mathrm{mass}}.
	\label{eq:total_loss}
\end{equation}

\subsection{\textbf{Training Procedure and Optimisation}}

Both stages are trained on four NVIDIA H100 80GB GPUs using PyTorch Distributed Data Parallel (DDP) with the NCCL backend. The Stage 2 UNet has approximately 256 million trainable parameters and is trained for up to 1000 epochs, with early stopping after 200 consecutive evaluation epochs without improvement in the held-out validation Pearson Correlation Coefficient on raw (non-log) precipitation. Optimisation uses AdamW (LR = $1e^{-4}$, weight decay $1e^{-3}$, gradient clipping at a global norm of 1.0) with a Cosine Annealing with Warm Restarts schedule $(T_{0} = 50$ epochs, minimum $\mathrm{LR} = 0.1 \times \mathrm{LR})$. Gradient accumulation over two steps yields an effective batch size of $16 \times 2 \times 4 = 128$. Mixed-precision (bfloat16) training is used for the forward pass and loss computation via PyTorch AMP, while parameter updates are kept in float32. An Exponential Moving Average (EMA) of the weights (decay 0.999) is applied during validation and inference, and validation metrics are synchronised across GPUs via all\_reduce.

To limit overfitting on the 49-year dataset, stochastic spatial augmentation is applied online random horizontal/vertical flips $(p = 0.5)$, $90^{\circ}$ rotations $(p = 0.5)$, brightness/contrast perturbation $(p = 0.4)$, additive Gaussian noise $(p = 0.25)$, and gamma adjustment $(p = 0.3)$  applied consistently to the precipitation, topographic, and D2M fields to preserve physical consistency, with coarse inputs re-derived from the augmented fine fields via average pooling.

\subsection{\textbf{Inference: Ensemble Sampling and Post-Processing}}
At inference, the trained Stage 2 UNet generates probabilistic downscaled fields through an ensemble of $N_{ens} = 8$ samples, each produced via a Denoising Diffusion Implicit Model (DDIM) sampling procedure over a 15-step EDM schedule. Starting from Gaussian noise at the maximum noise level, the sampler iteratively computes the denoised estimate using the EDM preconditioning coefficients and projects forward to the next noise level via the DDIM update rule:

\begin{equation}
	x_{\sigma_{i-1}}  	=  	x_{\sigma_i} 	+ 	\left(\sigma_{i-1} - \sigma_i\right)
	D_{\theta}\left(x_{\sigma_i}; \sigma_i\right)
	\label{eq:denoising_step}
\end{equation}
 The definition of each variable is given in the table \ref{tab:variables_eq7}.

\begin{table}[htbp]
	\centering
	\caption{Definition of the variables used in equation \ref{eq:denoising_step}.}
	\label{tab:variables_eq7}
	\begin{tabular}{|p{0.10\textwidth}|p{0.81\textwidth}|}
		\hline
		\textbf{Variable} & \textbf{Definition} \\
		\hline
		
		$x_{\sigma_i}$ &
		The noisy sample at noise level $\sigma_i$ --- the current state in the reverse diffusion (sampling) process at step $i$. \\
		\hline
		
		$x_{\sigma_{i-1}}$ &
		The updated (less noisy) sample at noise level $\sigma_{i-1}$, produced after one sampling step. \\
		\hline
		
		$\sigma_i$, $\sigma_{i-1}$ &
		The noise levels (standard deviations) at steps $i$ and $i-1$ in the discretized noise schedule; $\sigma_{i-1} < \sigma_i$ since sampling proceeds from high noise toward zero noise. \\
		\hline
		
		$D_{\theta}(x_{\sigma_i};\sigma_i)$ &
		The denoiser network (parameterized by trainable weights $\theta$), which takes the noisy sample $x_{\sigma_i}$ and its noise level $\sigma_i$ as input and outputs an estimate used to update the sample toward the clean signal. \\
		\hline
		
		$(\sigma_{i-1}-\sigma_i)$ &
		The step size --- the change in noise level between consecutive sampling steps. Negative in value, scaling how far the update moves along the direction given by $D_{\theta}$. \\
		\hline
		
	\end{tabular}
\end{table}

This deterministic trajectory enables efficient, consistent sample generation without the stochasticity of ancestral DDPM sampling; the final downscaled precipitation is obtained by adding the sampled residual to the Stage 1 mean prediction. Each sample then passes through a physics constraint layer (PhysicsGuide) enforcing two conservation properties: a dry-day mask, derived from the spatial mean of the coarse input, zeroes precipitation in globally dry regions, and an area-mean mass conservation constraint rescales the fine-resolution field so its spatial mean matches the upsampled coarse-resolution mean — bounded between 0.1 and 10.0 to avoid degenerate corrections.
We evaluate model performance every five training epochs using a set of complementary metrics on the held-out validation data. The main metric is the Pearson Correlation Coefficient (PCC) on raw precipitation values, calculated in two forms, a log-space weighted version (wPCC) and a raw-space version (R-PCC), and we use R-PCC to select the best-performing checkpoint. Three additional metrics help build a fuller picture of model quality. The Continuous Ranked Probability Score (CRPS), calculated from two ensemble members, measures how well-calibrated and sharp the predicted probabilities are. The Power Spectral Density tail ratio (PSD\_r) compares high-frequency spatial detail between predicted and observed fields, showing whether the model preserves fine-scale variability rather than smoothing it out. The Fractions Skill Score (FSS), calculated at a 0.5 mm/day rainfall threshold within a five-cell neighbourhood, checks how accurately the model captures the location of rainfall while allowing for small positional shifts. Together, these four metrics cover accuracy, spatial detail, probabilistic reliability, and the model's ability to capture extreme events.

\section{Results and Discussion}
The precipitation resulted from SRGAN as discussed in Section 3, exhibited a drawback of underestimating mean rainfall values. Moreover, the original CorrDiff model also continued to underestimate rainfall at the lower end of the precipitation range. To address this limitation, we developed a novel AI model using a hybrid Diffusion–UNet architecture based on the original CorrDiff framework and named as PrecipCorrDiff. This model was trained to generate 4x resolution precipitation data using the full dataset spanning 1975–2023, split into 80\% for training and validation and 20\% held out for testing. Model performance was evaluated on the test period (2014–2023) and compared against rain-gauge station observations for 2014–2018, the period for which station data were available.

\begin{figure}[ht]
	\center
	\includegraphics[width=0.99\textwidth,height=0.27\textheight]{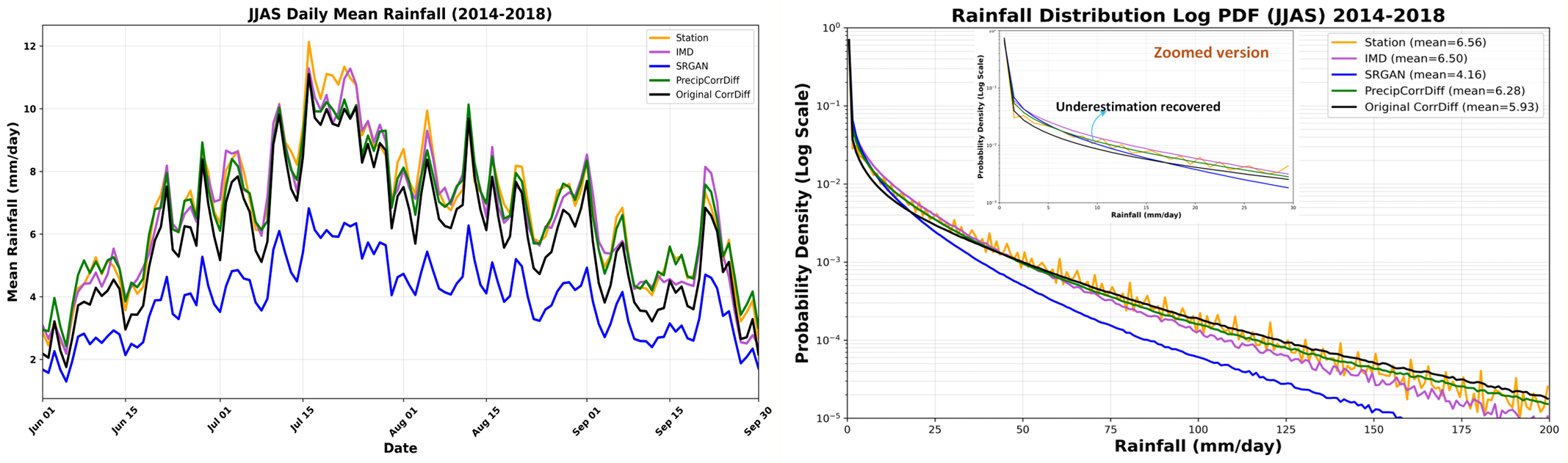}
	\caption{Comparison of rainfall representations, similar to Figure \ref{fig:srgan_underpredicted}, but including additional output from the Diffusion model. The left panel shows the mean rainfall density for JJAS months over the 2014–2018 period, while the right panel presents the corresponding raw data frequencies for the same period.}
	\label{fig:srgan_diffusion_distribution}
\end{figure}

The analysis focused on the monsoon season—June, July, August, and September (JJAS)—a 122-day period that represents the most critical window for Indian monsoon rainfall. We first compared the mean rainfall distributions, following the same approach used for Figure \ref{fig:srgan_underpredicted} in Section 2. This comparison, shown in Figure \ref{fig:srgan_diffusion_distribution}, proved that the proposed PrecipCorrDiff model exhibits a superior ability to capture the high-frequency fluctuations and peak intensities present in the station observations, thereby overcoming the underestimation limitation associated with the SRGAN approach.

Therefore, the model performance was evaluated using several other metrics which are mentioned in the method sections.  A comparison of the proposed Precipitation Corrector Diffusion (PrecipCorrDiff) model and the SRGAN method using four evaluation metrics are shown in figure 6. The left panels (a), (c), (e), and (g) depict the spatial distributions of PCC, RMSE, MASE, and PBIAS, respectively, whereas the right panels (b), (d), (f), and (h) show their corresponding temporal variations. 
The performance metrics were computed by comparing the model outputs with rain-gauge station observations, which were considered the ground truth (GT). The correlation coefficient (CC), calculated for both spatial and temporal distributions, is consistently higher for the PrecipCorrDiff model than for the SRGAN model (Figures 7(a) and 7(b)), indicating that PrecipCorrDiff reproduces the observed precipitation patterns more accurately. The spatial RMSE (Figure 7(c)) is comparable for both models, whereas the temporal RMSE (Figure 7(d)) is lower for PrecipCorrDiff, demonstrating its improved capability to capture the temporal variability of precipitation. A similar trend is observed for the mean absolute error (MAE), with PrecipCorrDiff yielding lower errors than SRGAN for both spatial and temporal analyses (Figures 7(e) and 7(f)). A more pronounced difference is evident in the percentage bias (PBIAS). PrecipCorrDiff exhibits consistently positive PBIAS values, whereas SRGAN shows predominantly negative PBIAS values for both spatial and temporal comparisons (Figures 7(g) and 7(h)). The negative PBIAS associated with SRGAN indicates a systematic underestimation of precipitation, particularly during high-rainfall events, whereas CorrDiff substantially reduces this bias.  Thus, the proposed PrecipCorrDiff model provides a more accurate and reliable representation of precipitation than the SRGAN model.

\begin{figure}[htpb]
	\center
	\includegraphics[width=0.81\textwidth,height=0.75\textheight]{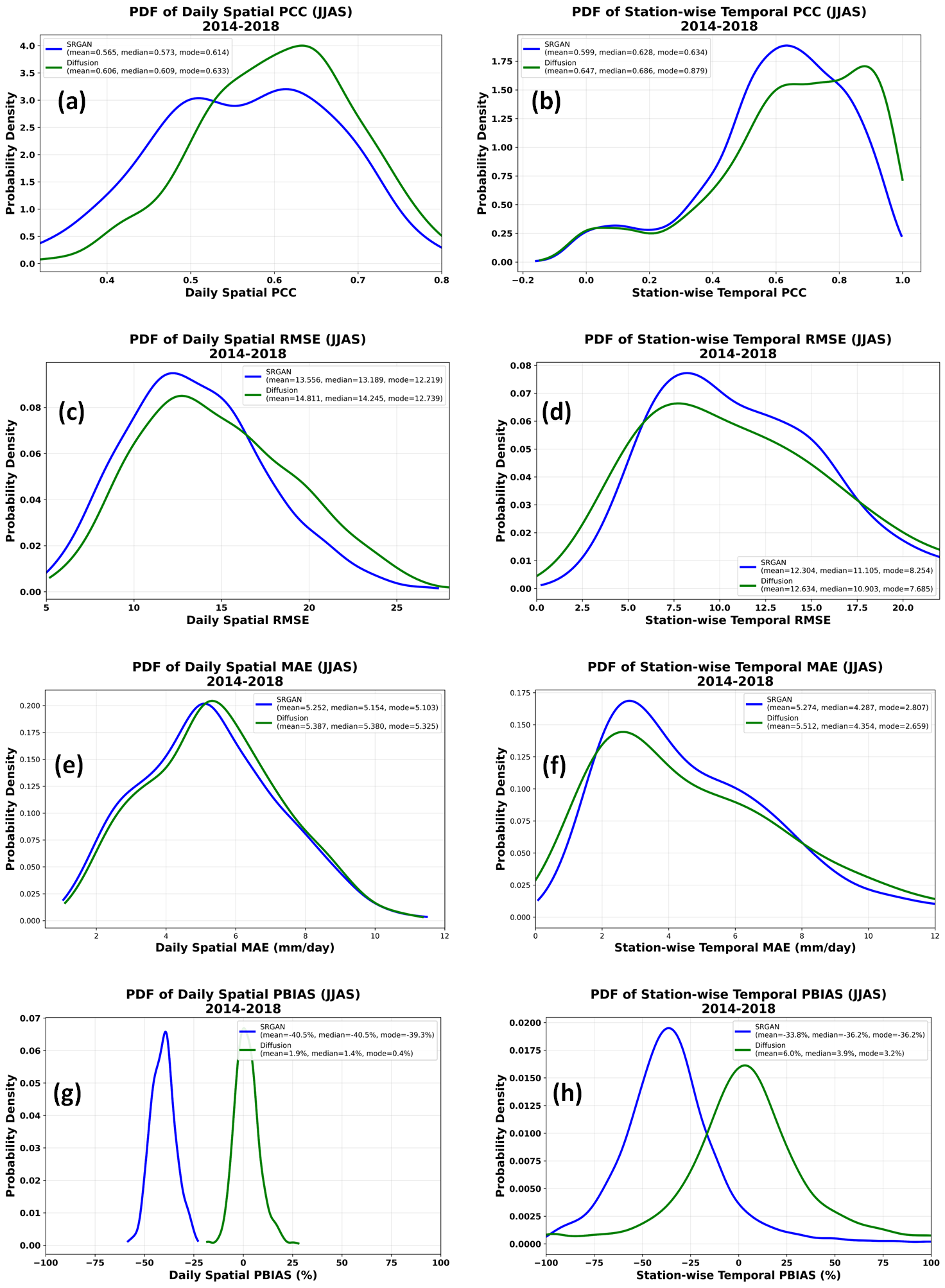}
	\caption{Performance comparison of the proposed Corrective Diffusion model and the SRGAN method using four evaluation metrics. The left panels (a), (c), (e), and (g) depict the spatial distributions of PCC, RMSE, MASE, and PBIAS, respectively, whereas the right panels (b), (d), (f), and (h) show their corresponding temporal variations. }
	\label{fig:anamoly_plots}
\end{figure}

Figure \ref{fig:anamoly_plots} compares the performance of the two models using anomaly- and error-based statistical metrics. To provide a more comprehensive evaluation, Figure 8 presents the distributions of several widely used skill-score metrics, including the Equitable Threat Score (ETS), Fractions Skill Score (FSS), Heidke Skill Score (HSS), Kling–Gupta Efficiency (KGE), and Threat Score (TS), computed by comparing the model outputs with the rain-gauge observations (GT). These metrics assess the ability of the models to detect precipitation events and reproduce the observed rainfall characteristics
As shown in Figure \ref{fig:anamoly_plots}(a), the distribution of ETS values is consistently shifted toward higher values for the PrecipCorrDiff model compared with SRGAN, with the median ETS increasing from 0.37 to 0.43, indicating improved event detection skill. The FSS distributions shown in Figure 8(b) also demonstrate superior performance of PrecipCorrDiff, with a median value of 0.86, compared with 0.83 for SRGAN, reflecting a better representation of the spatial organization of precipitation.

\begin{figure}[ht]
	\center
	\includegraphics[width=0.81\textwidth,height=0.55\textheight]{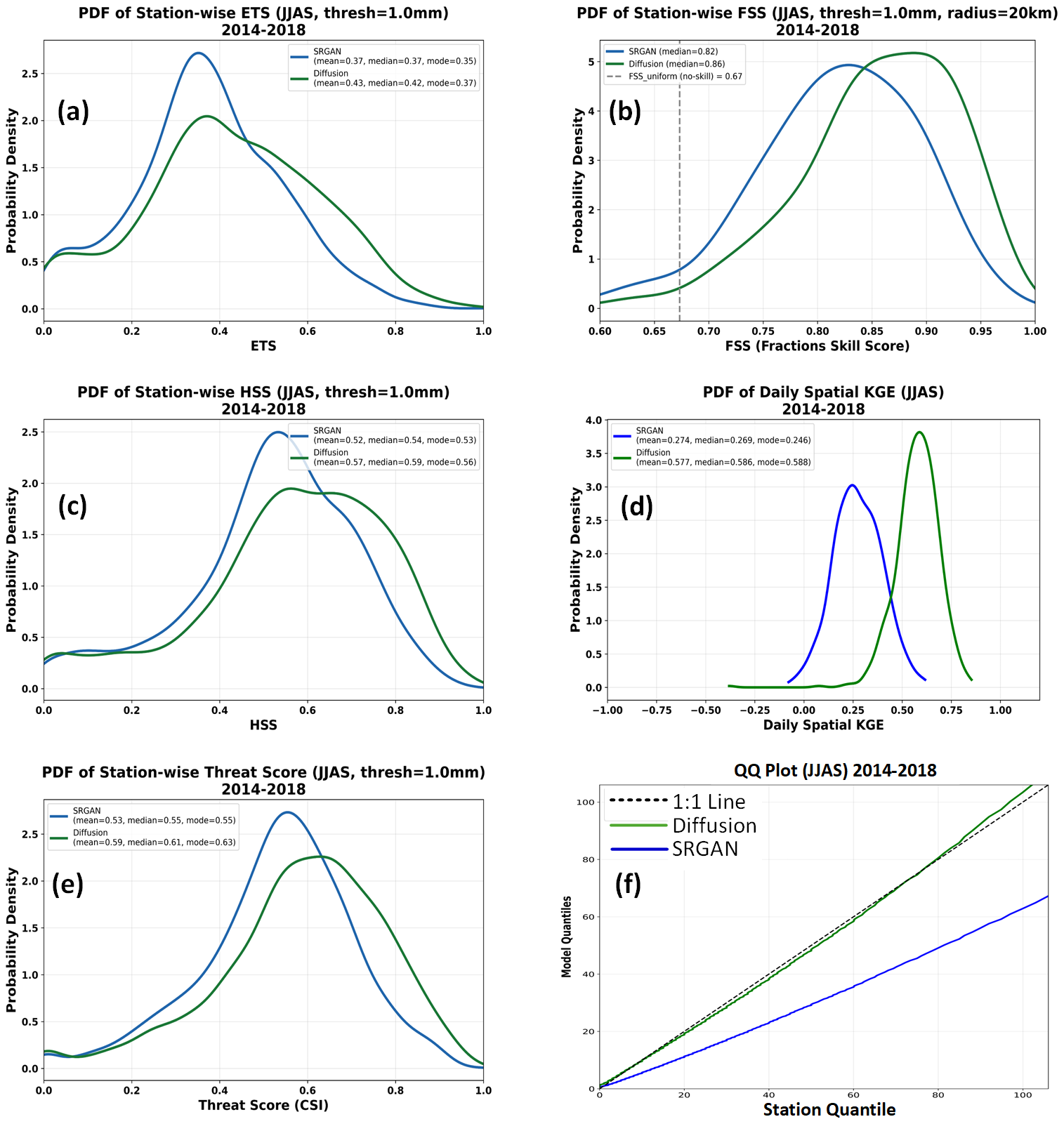}
	\caption{Comparison of the proposed Corrective Diffusion model and the SRGAN method based on five skill metrics and a quantile–quantile (Q–Q) plot. Panels (a)–(e) display the PDFs for ETS, FSS, HSS, KGE, and Threat Score (TS), respectively, while panel (f) presents the Q–Q plot comparing the precipitation distributions of the two models. All evaluation metrics were computed with station observations as the reference (ground truth). }
	\label{fig:skill_scores}
\end{figure}

Similar improvements are evident for the remaining skill scores presented in Figures \ref{fig:skill_scores}(c)–\ref{fig:skill_scores}(e). For HSS, KGE, and TS, the PrecipCorrDiff model consistently exhibits higher median values and distributions shifted toward larger scores than those of SRGAN. The largest improvement is observed for KGE, where the median increases from 0.269 for SRGAN to 0.586 for PrecipCorrDiff, indicating a substantially better representation of precipitation variability, correlation, and magnitude.  The quantile–quantile (Q–Q) plot in Figure \ref{fig:skill_scores}(f) further supports these findings. The PrecipCorrDiff estimates closely follow the 1:1 reference line over the entire range of precipitation intensities, whereas the SRGAN estimates increasingly deviate below the reference line at higher rainfall amounts, confirming the tendency of SRGAN to underestimate intense precipitation events. Overall, the skill-score analysis presented in Figure 8 corroborates the results obtained from the anomaly-based metrics in Figure 7, demonstrating the superior performance and robustness of the proposed PrecipCorrDiff model.

The analyses presented in Figures \ref{fig:anamoly_plots} and \ref{fig:skill_scores} were performed over the entire Indian landmass. However, rainfall characteristics exhibit substantial spatial variability across India due to differences in topography, circulation patterns, and monsoon dynamics. Therefore, to assess the robustness of the proposed model under diverse climatic conditions, a regional evaluation was carried out over the four homogeneous rainfall regions defined by the India Meteorological Department (IMD), namely Northern India, Northeastern India, Central India, and South Peninsular India. Figure 9 summarizes the regional comparison of the proposed PrecipCorrDiff model and the SRGAN model using both anomaly-based and skill-score metrics. The left panels present the anomaly-based metrics, including the Pearson Correlation Coefficient (PCC), Root Mean Square Error (RMSE), Mean Absolute Error (MAE), and Percentage Bias (PBIAS), while the remaining right panels show the corresponding skill-score metrics. The pie charts illustrate the station-wise comparison between the two models for each evaluation metric. Each chart is partitioned into two colored sectors, where the blue sector corresponds to the percentage of rain-gauge stations at which the SRGAN model performs better than PrecipCorrDiff, and the green sector represents the percentage of stations where the proposed PrecipCorrDiff model exhibits superior performance. Consequently, the relative area of each sector directly indicates the fraction of stations at which each model provides the better estimate.

\begin{figure}[htpb]
	\center
	\includegraphics[width=0.66\textwidth,height=0.79\textheight]{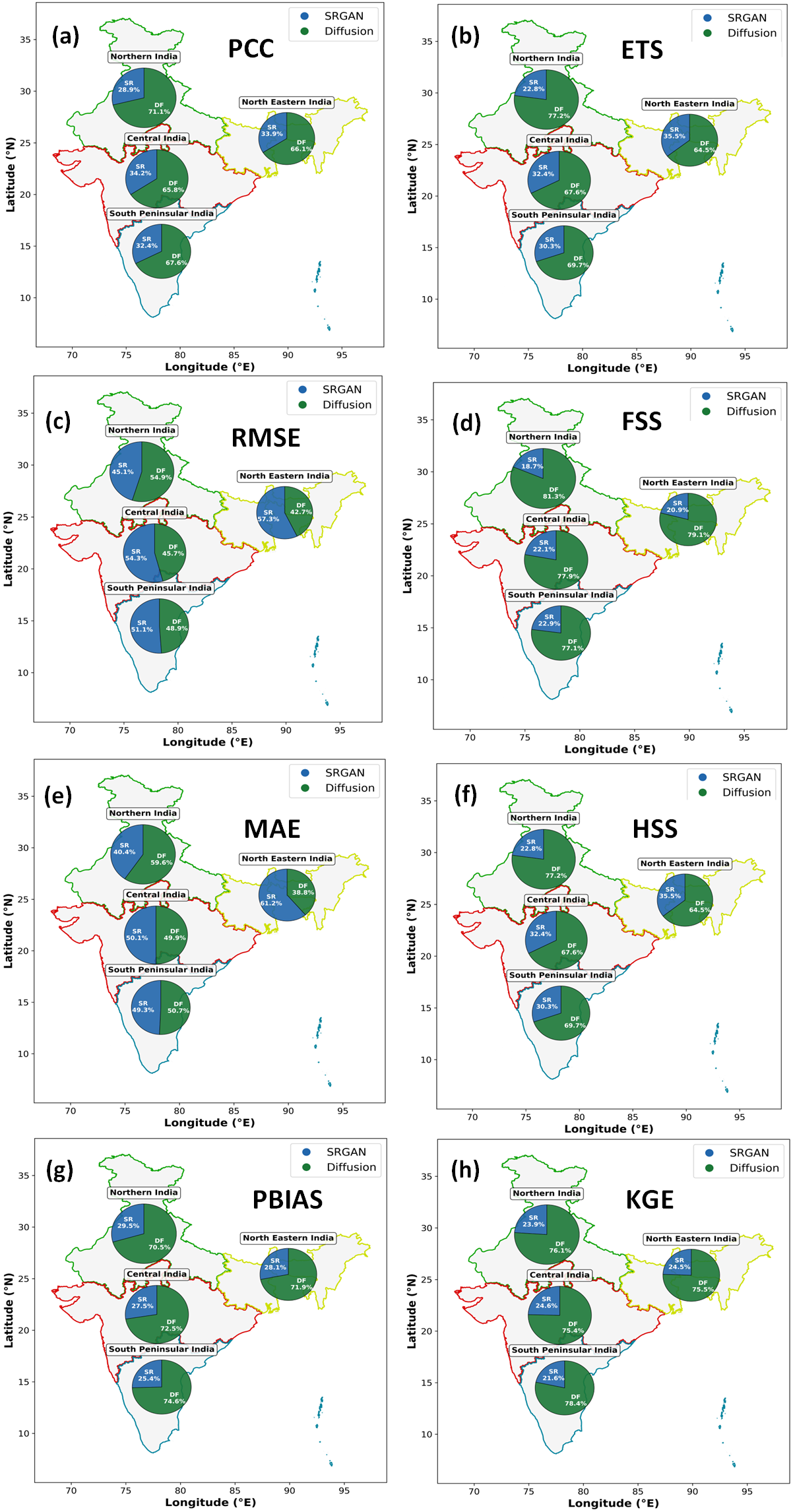}
	\caption{Area-wise performance comparison of model outputs against station data, based on selected evaluation metrics, shown using pie diagrams. The green portion of each pie chart represents the percentage of areas where the Diffusion model (PrecipCorrDiff) performed better, while the blue portion represents the percentage where SRGAN performed better. }
	\label{fig:pie_chart_4_regions_metric}
\end{figure}

As an example, Figure \ref{fig:pie_chart_4_regions_metric}(a) presents the station-wise comparison based on the PCC metric. The PrecipCorrDiff model achieves higher correlation than SRGAN at 71.1\% of stations in Northern India, 66.1\% in Northeastern India, 65.8\% in Central India, and 67.6\% in South Peninsular India. These results indicate that the proposed PrecipCorrDiff model consistently provides a better representation of the spatial and temporal variability of precipitation across all major monsoon regions of India. Figure \ref{fig:pie_chart_4_regions_metric}(c) presents the station-wise comparison based on the RMSE metric. The PrecipCorrDiff model outperforms SRGAN in Northern India, Central India, and South Peninsular India, whereas the improvement over Northeastern India is comparatively limited. In contrast, the proposed PrecipCorrDiff model consistently demonstrates superior performance for both MAE and PBIAS across all four regions, as illustrated in Figures \ref{fig:pie_chart_4_regions_metric}(e) and \ref{fig:pie_chart_4_regions_metric}(g). These results indicate that PrecipCorrDiff not only reduces the magnitude of prediction errors but also effectively mitigates the systematic bias observed in the SRGAN model.
	
The right-hand panels of Figure \ref{fig:pie_chart_4_regions_metric} (Figures \ref{fig:pie_chart_4_regions_metric}(b), \ref{fig:pie_chart_4_regions_metric}(d), \ref{fig:pie_chart_4_regions_metric}(f), and \ref{fig:pie_chart_4_regions_metric}(h)) present the regional comparison of the skill-score metrics using the same station-wise pie-chart representation. For each metric, the pie charts indicate the percentage of rain-gauge stations where either the PrecipCorrDiff (DF) or the SRGAN model performs better relative to the ground-truth observations. Across all four climatic regions, the PrecipCorrDiff model consistently outperforms SRGAN, with a larger proportion of stations exhibiting higher skill scores. These regional results confirm that the improvements achieved by the proposed PrecipCorrDiff model are robust and consistent across the diverse rainfall regimes of India.

\begin{figure}[ht]
	\center
	\includegraphics[width=0.95\textwidth,height=0.35\textheight]{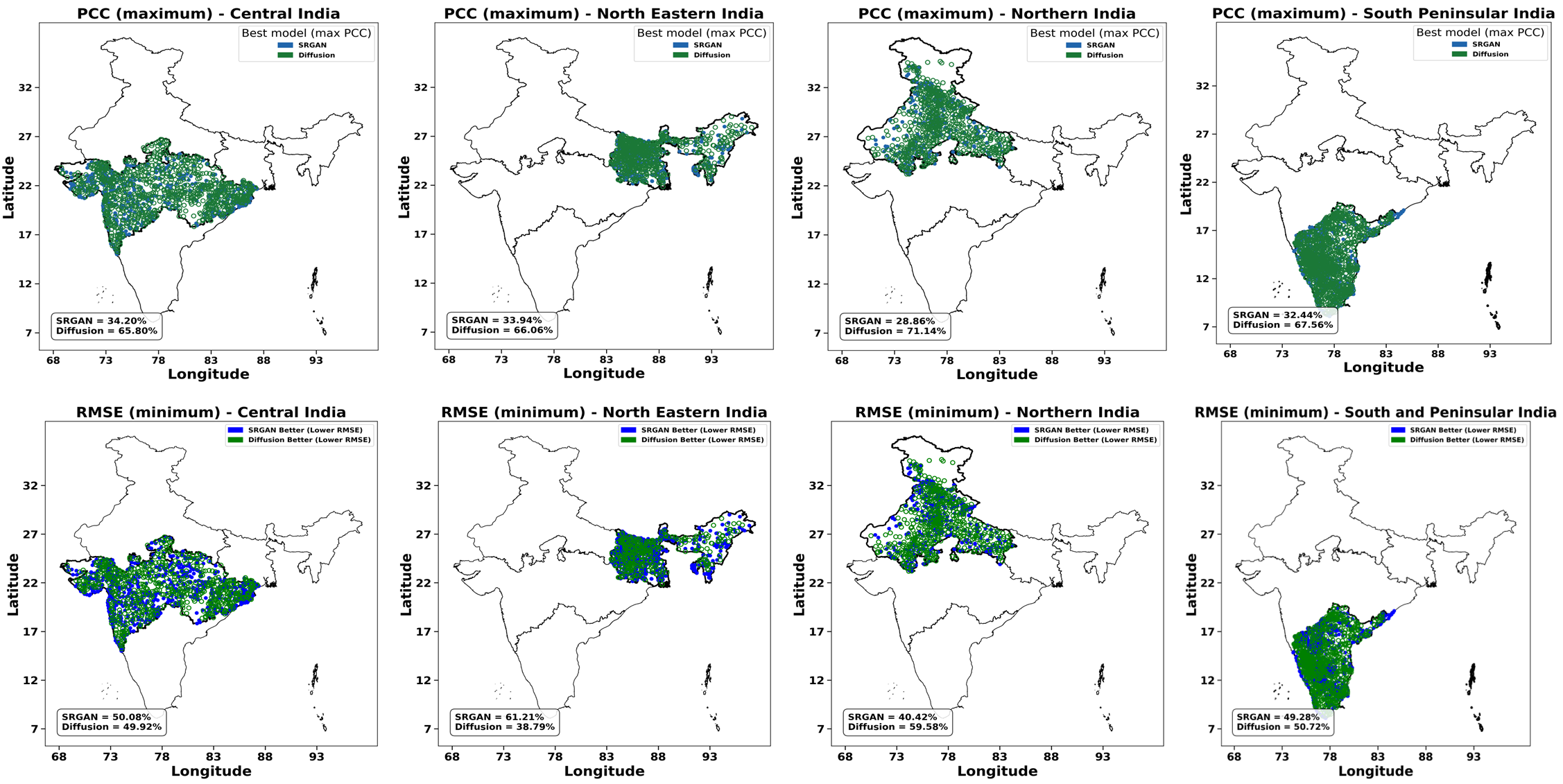}
	\caption{Area-wise model performance at each station using PCC and RMSE across four regions. The upper panels, from left to right, show station-wise PCC for Central India, Northeastern India, Northern India, and South Peninsular India. The corresponding station-wise RMSE values, in the same regional order, are shown in the lower panels. Green circles indicate locations where the Diffusion model performed better, and blue circles indicate locations where SRGAN performed better, with the total number of locations for each model also given as a percentage. }
	\label{fig:area_wise_performance}
\end{figure}

Similar to Figure \ref{fig:pie_chart_4_regions_metric}, the performance of both models is evaluated using two metrics, PCC and RMSE, at each station as the scatter plots are presented in Figure \ref{fig:area_wise_performance}. The maximum PCC and minimum RMSE values at each station location across the four regions are shown, along with the percentage of total stations won by each model annotated in the panel text. In all four regions, the maximum PCC was found to be higher for the Diffusion model than for SRGAN. This figure clearly shows that the Diffusion model outperformed SRGAN at more than 60\% of station locations based on PCC, while the RMSE-based comparison was considerably closer, with station-level wins split close to evenly between the two models in most regions.

\begin{figure}[ht]
	\center
	\includegraphics[width=0.95\textwidth,height=0.21\textheight]{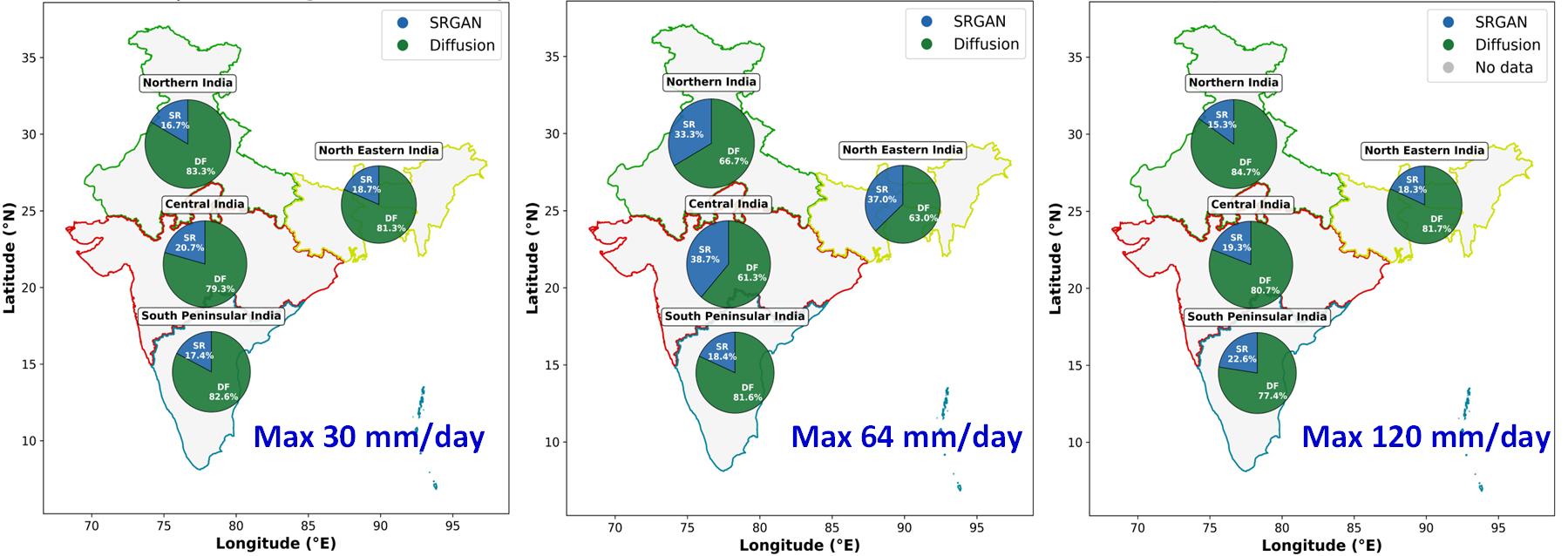}
	\caption{Performance comparison of the proposed Corrective Diffusion model and the SRGAN method across three rainfall categories based on RMSE in the four study regions. }
	\label{fig:category_wise_performance}
\end{figure}

All the analysis presented till now was based on all the rainfall values. However, one of the main issues in previous studies on downscaling remain capturing the extreme values by downscaling methods applied. In view of this, performance evaluation of the models has also been carried based rainfall categories which are referred as moderate rain (30mm/day),  heavy rain (64 mm/day) and very heavy rain (120 mm/day). 
\begin{figure}[ht]
	\center
	\includegraphics[width=0.90\textwidth,height=0.60\textheight]{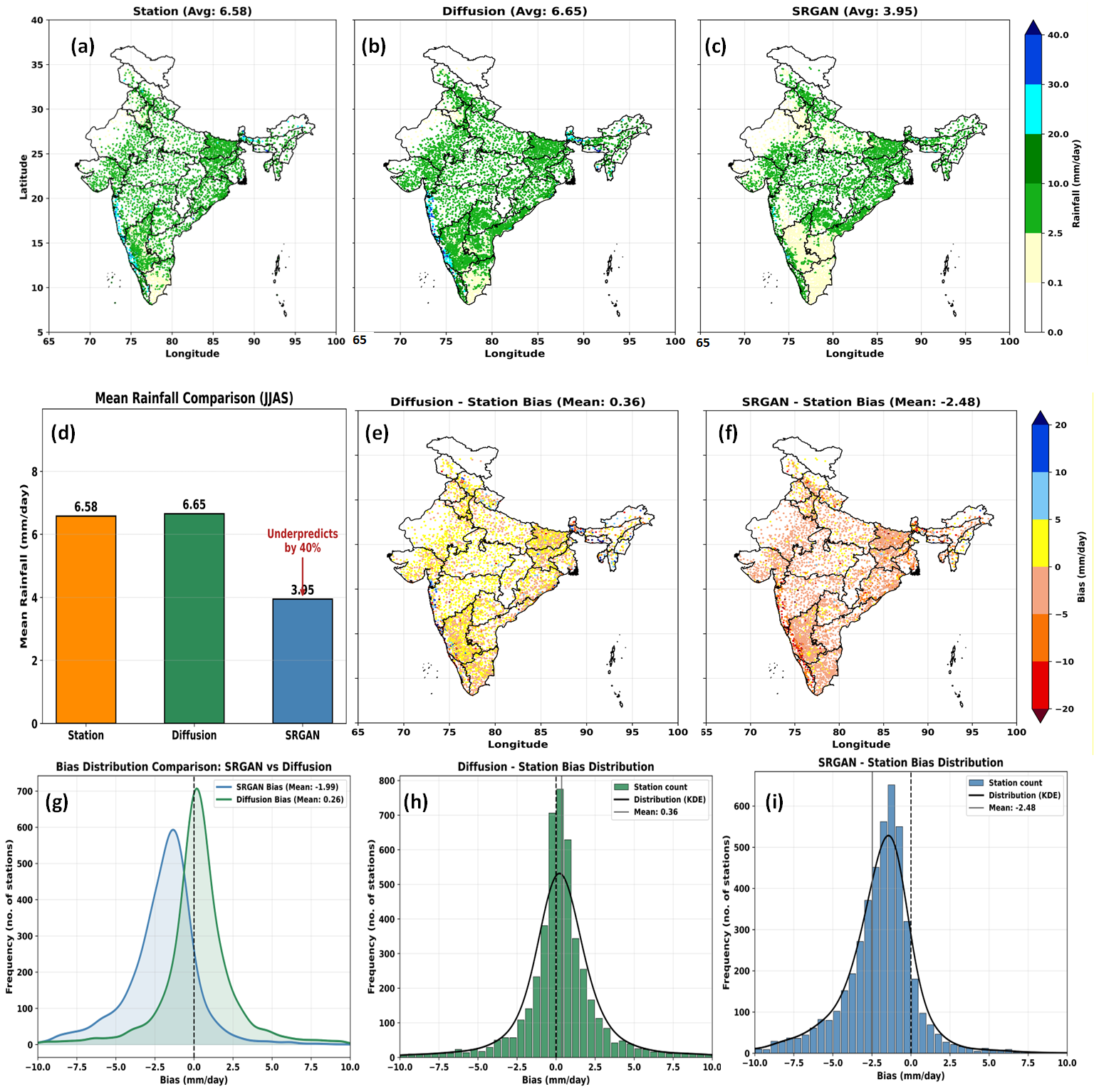}
	\caption{Station-based comparison of precipitation climatology between ground-truth observations and model outputs at $0.0625^o$ resolution. The model outputs were interpolated to the corresponding station locations for the test period (2014–2018). }
	\label{fig:mean_bias}
\end{figure}
Figure \ref{fig:category_wise_performance} presents performance of both models in four regions defined above using Pie chart cover area with same color sector notation as illustrated in figure \ref{fig:pie_chart_4_regions_metric}. The figure indicated that in all three categories, PrecipCorrDiff  method highly outperformed the SRGAN method specially for the very heavy rainfall category shown in the right most panels of figure \ref{fig:area_wise_performance}, the DF method outperformed more than 80\%  station location in all the four regions except the south peninsular region where it has the performance slightly less than 80\%. Same performance is recorded for moderate rainfall category (Max 30mm/day) where PrecipCorrDiff  method outperformed about 80\% or more station locations.

The mean precipitation underestimation associated with the SRGAN method, introduced earlier in the introduction section, is effectively addressed by the proposed PrecipCorrDiff framework. Beyond the underestimation pattern illustrated in Figure \ref{fig:srgan_diffusion_distribution}, we further evaluated the models by interpolating their outputs to station locations and comparing the resulting mean precipitation values with observed station data.  Figure \ref{fig:mean_bias} summarizes these comparisons: panel (a) presents the observed station precipitation, panel (b) shows the diffusion-based (PrecipCorrDiff) output interpolated to the station locations, and panel (c) shows the corresponding interpolated SRGAN output. The SRGAN results exhibit a clear underestimation of rainfall. The spatiotemporally averaged precipitation across all station locations is 3.95 mm, substantially lower than the observed mean of 6.58 mm. In contrast, the diffusion-based (PrecipCorrDiff) model produces an average precipitation of 6.65 mm, closely matching the observations. This difference is further illustrated in the histogram shown in Figure \ref{fig:mean_bias}(d), which indicates that SRGAN underpredicts rainfall by approximately 40\% relative to the observed values. The spatial distribution of prediction bias, defined as Predicted - Ground Truth, is shown in Figures 12(e) and 12(f) for the diffusion-based and SRGAN methods, respectively. As evident in Figure \ref{fig:mean_bias}(f), SRGAN exhibits a pronounced negative bias across nearly all station locations, whereas the diffusion-based model displays only minimal bias throughout the domain. 

\begin{figure}[htpb]
	\center
	\includegraphics[width=0.95\textwidth,height=0.85\textheight]{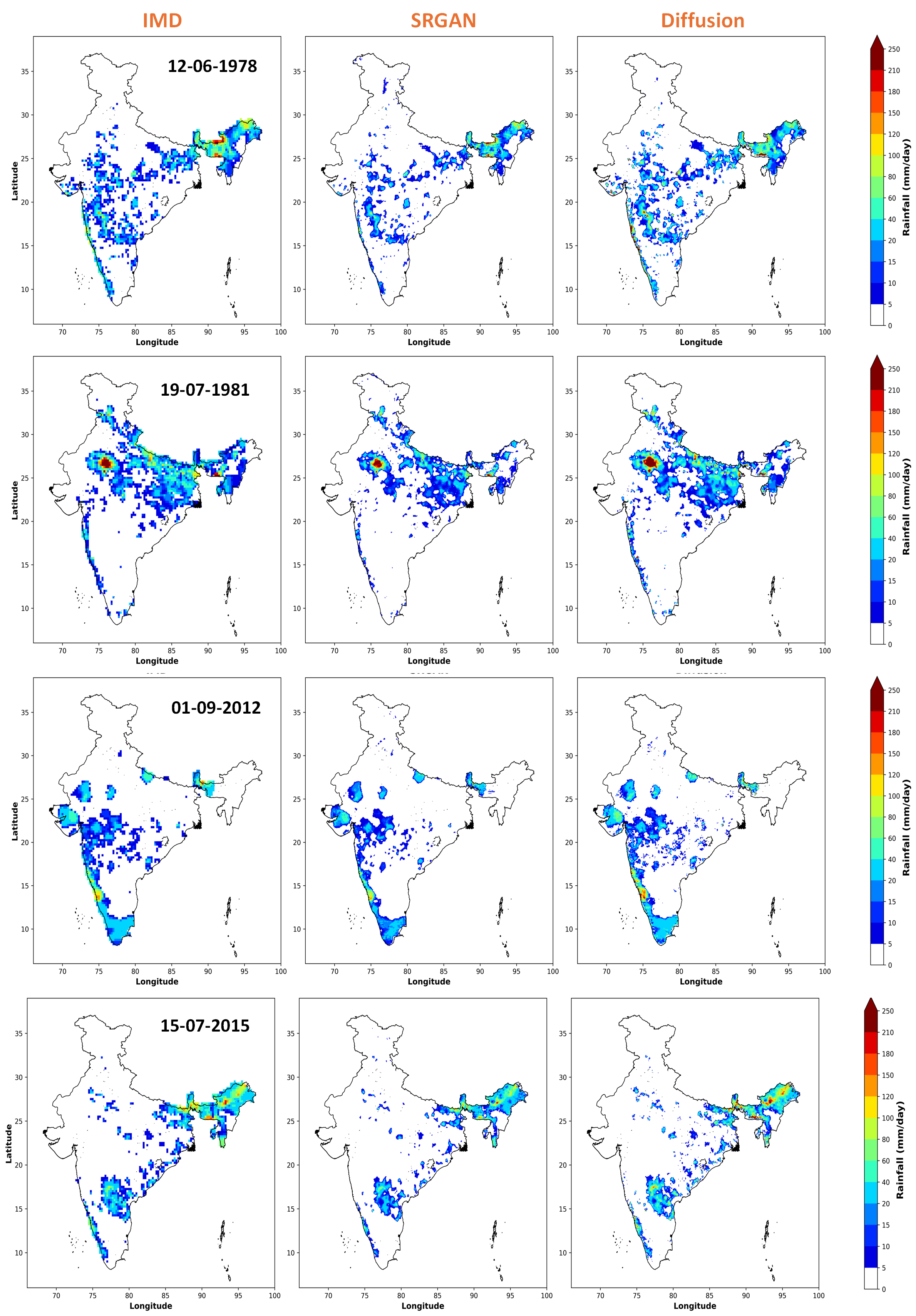}
	\caption{Comparison of the spatial distribution of rainfall from station observations, the SRGAN model, and the Diffusion model, demonstrating that the Diffusion model more accurately reproduces the observed spatial variability of rainfall. }
	\label{fig:map_plot}
\end{figure}

To further quantify these differences, we analyzed the distribution of station-wise bias values using probability density functions (PDFs) and histograms. Figure \ref{fig:mean_bias}(g) compares the PDFs of the biases for both methods, demonstrating that SRGAN biases are consistently shifted toward negative values, while the diffusion-model biases are centered close to zero. This behavior is further reinforced by the corresponding histograms in Figures \ref{fig:mean_bias}(h) and \ref{fig:mean_bias}(i), which show that the diffusion-based approach produces a nearly unbiased distribution, whereas SRGAN exhibits a systematic negative bias across the station network.

Apart from comparing the regional mean rainfall and the distribution of mean bias, we further evaluated the ability of the models to reproduce the spatial distribution of rainfall over India using four representative rainfall events. Figure 13 presents the observed and model-generated rainfall fields for 12 June 1978, 19 July 1981, 1 September 2012, and 15 July 2015, which represent distinct rainfall regimes over different parts of the country as well as different phases of monsoon annual cycles as depicted in figure 6. The 12 June 1978 event is characterized by intense rainfall over the Western Ghats and the northeastern region. The Diffusion model successfully captures both the location and intensity of these rainfall maxima, closely resembling the observed spatial pattern. In contrast, the SRGAN model produces a smoother rainfall field and underestimates the magnitude and spatial extent of the heavy precipitation. The 19 July 1981 event exhibits widespread rainfall over central India, extending across a large part of the monsoon core region. While both models reproduce the general rainfall belt, the Diffusion model provides a much better representation of the observed rainfall distribution and localized high-intensity regions, whereas SRGAN yields comparatively weaker and more diffuse rainfall patterns. The 1 September 2012 case is dominated by heavy rainfall along the Western Ghats, associated with strong orographic precipitation. The Diffusion model accurately preserves the sharp rainfall gradients and localized rainfall maxima over the windward side of the mountain range, whereas SRGAN tends to smooth these features and underestimate the peak rainfall intensity. Similarly, the 15 July 2015 event is characterized by intense precipitation over the northeastern states, including Sikkim, where the complex terrain leads to highly localized rainfall. The Diffusion model reproduces these fine-scale spatial features with much greater fidelity, while the SRGAN model fails to capture the observed intensity and spatial variability. Across all four representative events, the Diffusion model consistently provides closer agreement with the observed rainfall distribution by preserving the location, intensity, and spatial extent of precipitation systems. These case studies complement the statistical analyses of mean rainfall and mean bias, providing compelling visual evidence that the Diffusion model offers a more realistic representation of the spatial structure of precipitation than the SRGAN model.

Based on the above evaluation, the diffusion-based PrecipCorrDiff framework was selected as the final downscaling approach due to its superior ability to reproduce observed precipitation statistics and substantially reduce the systematic underestimation exhibited by SRGAN. Using the trained PrecipCorrDiff model, we generated a high-resolution precipitation dataset for the entire study period (1975–2023) by downscaling the original $0.25^o$ gridded data to a spatial resolution of $0.0625^o$. To facilitate reproducibility and enable independent evaluation of the model performance, the test dataset corresponding to the period 2014–2018 has been made publicly available through Zenodo. This open-access dataset allows users to directly compare the model output with observations and serves as a benchmark for future studies on precipitation downscaling and related applications.

\section{Conclusion}
Precipitation downscaling could provide high-resolution rainfall information required for regional-scale applications. In this study, we developed a deep learning framework based on the Corrector Diffusion (CorrDiff) approach, combined with a custom loss function, for downscaling the India Meteorological Department (IMD) gridded precipitation data. The proposed framework, termed the Precipitation Corrector Diffusion (PrecipCorrDiff),  built by modifying the original corrector diffusion based architecture, effectively addresses the limitations of the previously developed SRGAN-based approach, particularly its systematic underestimation of mean precipitation and inability to accurately reproduce the observed rainfall distribution.

A comprehensive evaluation of the SRGAN and PrecipCorrDiff downscaled products was performed against station observations using multiple statistical metrics, spatial mean precipitation, rainfall distributions, mean bias, and bias distributions. Across all evaluation measures, PrecipCorrDiff consistently outperformed the SRGAN model. The proposed framework successfully reproduced the observed mean precipitation, eliminating the systematic underestimation present in the SRGAN outputs while maintaining close agreement with both the original $0.25^o$ IMD dataset and station observations. Furthermore, PrecipCorrDiff demonstrated an improved capability to capture extreme rainfall events across all four climatologically distinct rainfall regions considered in this study. The spatial bias of PrecipCorrDiff, largely, remained close to zero throughout the domain, in contrast to the pronounced negative bias exhibited by the SRGAN method.

Based on its consistently superior performance, the PrecipCorrDiff framework was employed to generate a high-resolution precipitation dataset by downscaling the IMD $0.25^o$ gridded precipitation data to a spatial resolution of 0.0625° for the period 1975–2023. This study makes several key contributions toward improving high-resolution precipitation downscaling over India:

\begin{enumerate}[itemsep=2pt, topsep=2pt, parsep=0pt, partopsep=0pt]
	\item The proposed framework successfully corrects the all-India area-averaged annual rainfall cycle, closely reproducing the observed seasonal progression of precipitation (Fig.~6a).
	
	\item It substantially improves the rainfall value bin distribution, including a marked correction in extreme rainfall values that were previously underestimated by the SRGAN-based approach (Fig.~6b).
	
	\item The model achieves reliable regional-scale accuracy in capturing extreme rainfall events across all four climatologically distinct monsoon regions of India (Fig.~9).
	
	\item Station-based climatological evaluation confirms a consistent improvement in reproducing observed rainfall characteristics at individual station locations, further validating the model's performance at the local scale (Fig.~12).
\end{enumerate}

To ensure reproducibility and enable independent evaluation, the downscaled test dataset covering the 2014–2018 period has been made publicly available through Zenodo \citep{Kumar2026PrecipCorrDiffData}. The generated high-resolution dataset has the potential to support the validation of IITM High-Resolution Global Forecast Model (also known  as the Bharat Forecast System), which operates at an approximate spatial resolution of 6 km \citep{Krishna2025} and similar model at this resolution. In addition, the dataset provides a valuable resource for developing next-generation AI-based precipitation forecasting models and for a wide range of hydrological, climate, and water-resource applications requiring high-resolution rainfall information.


 \section*{Acknowledgments}
IITM Pune is financially supported by the Ministry of Earth Sciences, Government of India. 
BK, RC,BBS and TP acknowledge the HPC facilities provided by IITM Pune.

%
%
\section*{DataData Availability Statement}
The downscaled precipitation dataset for the 2014–2018 test period is publicly available through Zenodo \citep{Kumar2026PrecipCorrDiffData}. The dataset can be accessed at https://zenodo.org/records/21819376.

\bibliographystyle{abbrvnat}
\bibliography{reference_PrecipCorrDiff}

\end{document}